\documentstyle[emulateapj,epsf,apjfonts]{article}

\begin{document}

\title{First Determination of the Distance and Fundamental Properties of
an Eclipsing Binary in The Andromeda Galaxy\altaffilmark{1,2}}
\altaffiltext{1}{Based on observations obtained at the Gemini Observatory,
which is operated by the Association of Universities for Research in
Astronomy, Inc., under a cooperative agreement with the NSF on behalf of
the Gemini partnership: the National Science Foundation (United States),
the Particle Physics and Astronomy Research Council (United Kingdom), the
National Research Council (Canada), CONICYT (Chile), the Australian
Research Council (Australia), CNPq (Brazil) and CONICET (Argentina)}
\altaffiltext{2}{Based on observations made with the Isaac Newton
Telescope operated on the island of La Palma by the Isaac Newton Group in
the Spanish Observatorio del Roque de los Muchachos of the Instituto de
Astrof\'{\i}sica de Canarias}

\author{Ignasi Ribas\altaffilmark{3,4}, Carme Jordi\altaffilmark{4,5},
Francesc Vilardell\altaffilmark{5}, Edward L. Fitzpatrick\altaffilmark{6},
Ron W. Hilditch\altaffilmark{7}, and Edward F. Guinan\altaffilmark{6}}

\altaffiltext{3}{Institut de Ci\`encies de l'Espai -- CSIC, Campus UAB,
Facultat de Ci\`encies, Torre C5 - parell - 2a, 08193 Bellaterra,
Spain; e-mail: iribas@ieec.uab.es}
\altaffiltext{4}{Institut d'Estudis Espacials de Catalunya (IEEC), Edif.
Nexus, C/Gran Capit\`a, 2-4, 08034 Barcelona, Spain}
\altaffiltext{5}{Dept. d'Astronomia i Meteorologia, Universitat de 
Barcelona, Avda. Diagonal 647, 08028 Barcelona, Spain; 
e-mail: carme,fvilarde@am.ub.es}
\altaffiltext{6}{Dept. of Astronomy \& Astrophysics, Villanova 
University, Villanova, PA 19085, USA; e-mail: fitz@astronomy.villanova.edu,
edward.guinan@villanova.edu} 
\altaffiltext{7}{School of Physics and Astronomy, University of St Andrews, 
North Haugh, St Andrews KY16 9SS, UK; e-mail: rwh@st-andrews.ac.uk}

\begin{abstract}
We present the first detailed spectroscopic and photometric analysis of an 
eclipsing binary in the Andromeda Galaxy (M31). This is a $19.3$-mag 
semi-detached system with components of late-O and early-B spectral types. 
From the light and radial velocity curves we have carried out an accurate 
determination of the masses and radii of the components. Their effective 
temperatures have been estimated from the modeling of the absorption line 
spectra. The analysis yields an essentially complete picture of the 
properties of the system, and hence an accurate distance determination to 
M31. The result is $d=772\pm44$ kpc ($(m-M)_{\circ}= 24.44\pm0.12$ mag). 
The study of additional systems, currently in progress, should reduce the 
uncertainty in M31 distance to better than 5\%.
\end{abstract}


\section{Introduction}

Accurate distance measurements to the Local Group galaxies are crucial to 
calibrating the Cosmic Distance Scale and to determining the age and 
evolution of the Universe. As major rungs on the cosmic distance ladder, 
these galaxies serve as calibrators for extragalactic distance indicators, 
such as Cepheid and RR Lyrae variables, novae, supernovae, globular 
clusters, etc, reaching far beyond the bounds of the Local Group (Hodge 
1981). Once a Local Group galaxy's distance is known, all of its various 
stellar populations are available as potential standard candles. The Large 
Magellanic Cloud (LMC) has traditionally been used because of its 
proximity to the Milky Way. However, its low metallicity, and irregular 
geometry has posed some difficulties as illustrated by the large spread in 
distances derived from different methods (e.g., Gibson 2000), although 
recent determinations seem to reach better agreement (Alves 2004). The 
Andromeda Galaxy (M31) is potentially a first-class distance calibrator 
(Clementini et al. 2001). Its main advantages are a simple geometry, a 
large and diverse stellar population, and a chemical composition and 
morphology similar to our Galaxy and galaxies used for distance estimation 
(Freedman et al. 2001). Also, M31 can provide an absolute calibration of 
the Tully-Fisher relationship.

The distance to M31 has been estimated using a variety of methods (e.g., 
Holland 1998; Walker 2003; McConnachie et al. 2005) with resulting 
distance moduli in the range $(m-M)_{\circ}$=24.0--24.5 mag. Eclipsing 
binaries (EBs) are excellent distance indicators because they yield 
fundamental determinations of the components' radii and luminositites 
(Guinan et al. 1998). EBs have been used to determine accurate distances 
to the LMC (e.g., Fitzpatrick et al. 2003; Ribas 2004) and SMC (e.g., 
Hilditch, Howarth, \& Harries 2005). In addition to providing accurate 
distances, the fundamental stellar properties from EBs, such as masses and 
radii, are of great value for studying the structure and evolution of 
stars formed in different environments.

The first discoveries of M31 EBs ($\sim$60 systems) came from photographic 
surveys (Baade \& Swope 1965 and references therein). More recently, the 
{\sc direct} group has reported about 90 new systems (see Bonanos et al. 
2003 and references therein). In 1999 we undertook or own wide-field CCD 
photometric survey (Ribas et al. 2004) to discover and measure multi-band 
EB light curves with sufficient accuracy for a reliable determination of 
their properties. From our survey we have selected several EBs for 
spectroscopic follow up. Here we present the first detailed photometric 
and spectroscopic analysis of an EB in M31 resulting in an accurate 
determination of its fundamental properties and distance.  This EB, 
designated M31VJ0044380+41292350, has an out-of-eclipse $V$ magnitude of 
19.3 and J2000.0 coordinates $\alpha=00^{\mathrm h} 44^{\mathrm m} 
38\fs0$, $\delta=+41\arcdeg 29\arcmin 23\farcs50$, and was reported by the 
{\sc direct} project under the identifier V12650 D31C (Stanek et al. 
1999).

\section{Observations and reductions}

Photometry in $B$ and $V$ passbands of a $34\arcmin\times34\arcmin$ field 
in the Northeastern quadrant of M31 was acquired with the Wide Field 
Camera (WFC) of the 2.5-m Isaac Newton Telescope. A total of 21 nights 
were allotted to our program over the course of four years (1999--2003) 
and $\sim$265 images in each filter were obtained. Photometry in the 
crowded spiral arms of M31 was performed using the Wo\'zniak (2000) 
implementation of the Difference Image Analysis (DIA) algorithm (Alard 
2000). The numerous parameters involved in the reductions were fine-tuned 
for best performance with the WFC images and modifications to the original 
code were made to improve some aspects. The resulting light curves have 
$\sim$240 photometric measurements in $V$ and $\sim$250 in $B$, with 
individual errors of about 0.01 mag at magnitude $V=19-20$. Full details 
of the reduction procedure and the resulting catalog will be given in a 
forthcoming publication (Vilardell, Ribas, \& Jordi 2005).

To complement our photometry we considered the discovery observations from 
the {\sc direct} group (Stanek et al. 1999) with a 1.3-m telescope. 
Photometry was obtained mostly in the $V$ (156 measurements) and $I$ (54 
measurements) bands, with average individual uncertainties of 0.04 mag and 
0.08 mag, respectively. As shown in \S \ref{sec:fits}, we used the {\sc 
direct} $V$ photometry to better constrain the orbital period and to check 
for consistency with our measurements (i.e., DIA vs. PSF photometry).

Spectroscopic observations were obtained with the 8-m Gemini-N telescope 
(program ID GN-2004B-Q-9). We used the GMOS spectrograph with a custom 
mask designed to deliver spectroscopy of a number of targets in a 
5\farcm5$\times$5\farcm5 field of view, including the EB in the present 
study. The instrument was set to the highest possible resolution of 
$R=3744$ ($\approx$80 km s$^{-1}$ per resolution element) using a slit 
width of 0\farcs5. The spectra cover a wavelength range of 3900--5350~\AA\ 
with a sampling of about 2.6 pixels per resolution element. Because of the 
instrument design, there are two gaps in the spectral dispersion direction 
in the wavelength intervals 4367--4379~\AA\ and 4858--4870~\AA, the latter 
affecting the H$\beta$ Balmer line. A total of 8 spectra with exposure 
times of 4100 s (2$\times$2050 s) were obtained, 4 of which were timed to 
cover both quadratures. The other 4 observations were taken at random 
phases and one of these happens to be also near quadrature. Reduction of 
the raw CCD frames was carried out with the {\sc iraf} Gemini package 
V1.7. The typical S/N of the resulting spectra is 15--30.

To derive radial velocities we considered several approaches based on both 
spectral disentangling and two-dimensional cross-correlation. We carried 
out numerous tests with the spectral disentangling codes {\sc tangle} 
(Harries, Hilditch, \& Howarth 2003) and {\sc korel} (Hadrava 1995), and 
considering different wavelenth intervals. Using a grid search method we 
found velocity semi-amplitudes in the ranges $K_1=175-200$~km~s$^{-1}$ and 
$K_2=260-300$~km~s$^{-1}$. The relatively large spread is a consequence of 
the noisy $\chi^{2}$ surface and it was not possible to establish a 
definitive solution for the values of $K_{1,2}$. Alternatively, we used 
the {\sc todcor} two-dimensional cross-correlation algorithm (Zucker \& 
Mazeh 1994). We calculated individual radial velocities for each spectrum 
by cross-correlating with synthetic templates from Kurucz ATLAS9 
models\footnote{http://kurucz.cfa.harvard.edu/} and the OSTAR2002 library 
(Lanz \& Hubeny 2003). We considered several spectral intervals and 
templates of different temperatures. The various runs yielded velocities 
in very good mutual agreement. The resultant radial velocities for the 
five spectra taken near quadratures are listed in Table \ref{tab:rvs}. The 
velocities obtained when masking out the Balmer H$\gamma$ and H$\delta$ 
lines, albeit of lower quality, are also in agreement.

\section{Modeling of the Light and Radial Velocity Curves}
\label{sec:fits}

The light and radial velocity curves were modeled using the 
Wilson-Devinney (W-D) code (Wilson \& Devinney 1971; Wilson 1990) in its 
2003 version. Initial tests indicated that the system is best described by 
a semi-detached configuration with the secondary star filling its Roche 
lobe. Also, those tests suggested that the light and radial velocity 
curves be modeled separately because of the very few radial velocity 
measurements available and the possibility of a bias arising in the mass 
ratio of this semi-detached system. Fitting was carried out iteratively 
until full consistency was achieved.

In the radial velocity fit, the adjustable parameters were the orbital 
semi-major axis ($a$), the mass ratio ($q$) and the systemic velocity 
($\gamma$). In the case of the light curves, we considered simultaneously 
our $B$ and $V$ observations, and the {\sc direct} $V$ light curve (the 
{\sc direct} $I$-band photometry was discarded because of the poor phase 
coverage and low quality). The light curve fits used time as the 
independent variable and thus the time of minimum ($T_{\rm min}$) and the 
orbital period ($P$) were adjusted. The rest of the adjustable parameters 
were the orbital inclination ($i$), the temperature of the secondary 
component (${T_{\rm eff}}_2$) -- the temperature of the primary was 
adopted from \S \ref{sec:teff}, -- the surface potential of the primary 
($\Omega_1$), and the luminosity of the primary in each passband ($L_1$). 
A circular orbit was adopted, and the gravity brightening coefficients and 
the bolometric albedos were set to unity in accordance with the radiative 
atmospheres of the components. The limb darkening coefficients 
(square-root law) were computed at each iteration from Kurucz ATLAS9 
atmosphere models.

\begin{figure}[!ht]
\epsscale{0.5}
\plotone{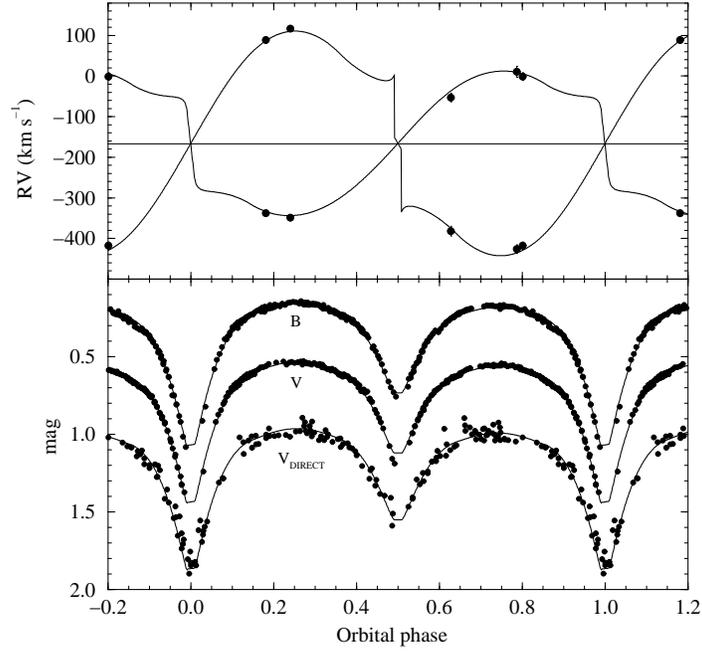}
\figcaption{Light and radial velocity curve fits. 
\label{fig:fits}}
\end{figure}

Convergence in the fits was reached rapidly and tests from different 
starting points indicated the uniqueness of the solution. It also became 
evident that the first quadrature is brighter than the other quadrature by 
about 0.025 mag in $B$ and 0.020 mag in $V$. This is frequently observed 
in semi-detached systems with active mass transfer and may arise from the 
impact of a stream of matter on the accretor. We modeled this effect in 
W-D by including a hot spot on the primary component. Assuming such 
circular spot to be 40\% hotter than the photosphere its predicted radius 
is $16\arcdeg\pm1\arcdeg$ and located at a longitude of 
$277\arcdeg\pm6\arcdeg$ (measured counter-clockwise from the line of star 
centers). Adding a spot in the model has little impact on the intrinsic 
properties of the system components but improves the quality of the fit in 
the out-of-eclipse region. The final r.m.s. residuals are 0.013 mag in 
$B$, 0.013 mag in $V$, and 0.046 mag in the {\sc direct} $V$ light curve. 
The residuals of the radial velocities are 5.2 km~s$^{-1}$ and 4.6 
km~s$^{-1}$ for the primary and secondary components, respectively. The 
resulting best-fitting elements are listed in Table \ref{tab:prop}, 
together with the fundamental stellar properties. The light and radial 
velocity curves with their respective fits superimposed are shown in Fig. 
\ref{fig:fits}. We also carried out an independent fit to the light curves 
with the {\sc light2} code (Hill \& Rucinski 1993), obtaining results 
almost identical to the W-D solution.

A possible concern with light curves measured using DIA photometry is the 
effect of an incorrect reference flux that may cause a bias in the scale 
of the light curves (Michalska \& Pigulski 2005). Careful tests were 
carried out to ensure that this was not the case. Also, the excellent 
agreement between the fits to our DIA photometry and the fits to the {\sc 
direct} PSF photometry strongly suggests that the flux zeropoint is 
correct. In addition, we ran light curve fits with a variable third light 
($L_3$) contribution. A non-zero value (either positive or negative) of 
$L_3$ might be indicative of problems with the flux scale. This did not 
occur and the solutions converged to $L_3\approx0$. Such result has 
another interesting consequence of ruling out possible blends with 
unresolved companions. The absence of positive $L_3$ is not unexpected 
since the light curves have the maximum possible depth ($i\sim90\arcdeg$) 
and preclude the existence of any additional light.

\section{Modeling of the Optical Spectra}
\label{sec:teff}

\begin{figure}[!ht]
\epsscale{1.0}
\plotone{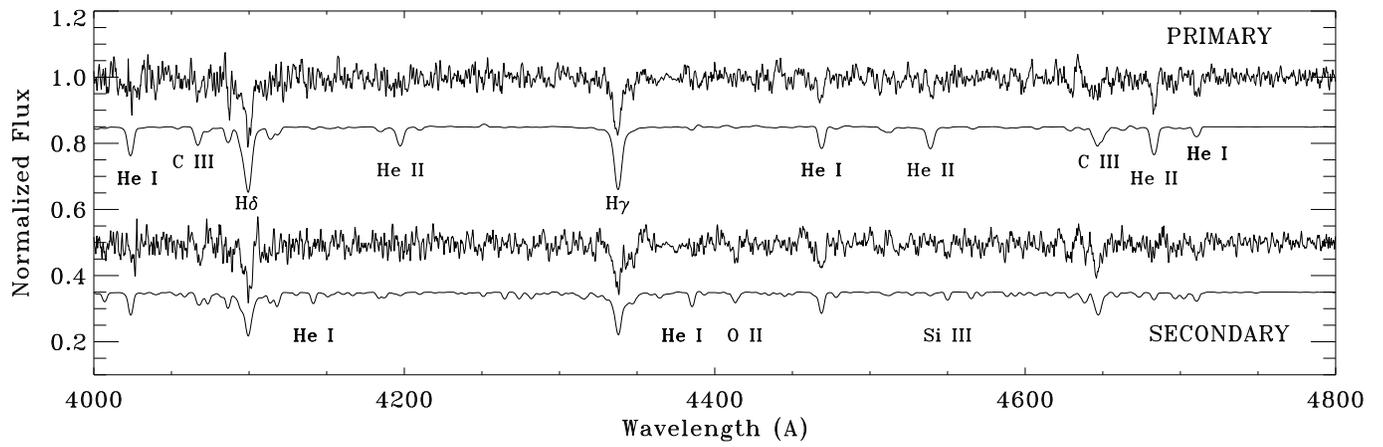}
\figcaption{Comparison of the individual ``disentangled'' spectra with 
NLTE synthetic spectra.
\label{fig:spec}}
\end{figure}

The photometric data available (i.e., $B$ and $V$) are, by themselves, 
insufficient to determine the temperatures and reddening of the system. 
However, valuable supplemental information can be derived from the optical 
spectra used for the radial velocity determinations. To access this, we 
ran the {\sc korel} program by fixing all the parameters to the orbital 
solution described in \S \ref{sec:fits}. This resulted in a 
``disentangled'' spectrum for each component of the system, with a S/N of 
about 40. These two normalized spectra are shown in Figure \ref{fig:spec}. 
We then modeled this pair of spectra simultaneously using the OSTAR2002 
grid of NLTE model atmospheres produced by Lanz \& Hubeny (2003). The fits 
were constrained by the temperature ratio, surface gravities, and 
brightness ratio determined from the binary analysis. We solved for the 
constrained temperatures, a single metallicity applying to both stars, and 
individual values of $v \sin i$. These values are listed in Table 
\ref{tab:prop} and the best-fitting models are shown in Figure 
\ref{fig:spec}, below each of the stellar spectra. Table \ref{tab:prop} 
also lists values of $M_V$ and $(B-V)_{\circ}$ for the system. These were 
produced by scaling the surfaces fluxes predicted by the OSTAR2002 models 
with the observed radii of the stars and performing synthetic photometry 
on the resultant energy distributions. The photometry was calibrated as 
described in Fitzpatrick \& Massa (2005).

The 1-$\sigma$ uncertainties in the results of the spectral analysis were 
determined by a Monte Carlo technique. First, we created simulations of 
the spectra by combining our pair of best-fit models with 50 different 
random noise realizations, corresponding to $S/N=40$. For each simulated 
spectrum pair, we generated a simulated set of binary parameters (e.g., 
${T_{\rm eff}}_2/{T_{\rm eff}}_1$), by combining the best-fit values with 
gaussian noise, based on the 1-$sigma$ uncertainties for each value. 
Finally, we fit each pair of simulated spectra as described above and 
adopted the standard deviations of the parameters among the 50 simulations 
as our uncertainties.

\section{Fundamental Properties and Distance}

This is the first pair of stars in M31 for which fundamental properties 
have ever been directly determined. The masses and radii in Table 
\ref{tab:prop} have relative uncertainties of 6--7\% and 2.5--3\%, 
respectively, which is remarkable given the faintness of the star. This 
opens the field of detailed investigations of stellar (single and binary) 
evolution models in this neighboring giant spiral galaxy.  But here we 
focus on one of the main drivers of our program, which is the 
determination of the distance to the EB system. In our studies in the LMC 
we have adopted an approach based on the fit of the spectral energy 
distribution to determine the temperature, reddening, and distance to the 
system. At this time, such method is not applicable to this M31 binary and 
we have employed a different approach that uses the available data.

The calculation of the distance is straightforward because the spectral 
analysis yields values for the absolute magnitudes $M_V$ of the components 
and also the combined $M_V$ of the system (see Table \ref{tab:prop}). The 
distance modulus follows directly from the equation: $(m - M)_{\circ} = V
- M_{\rm V} - A_V$. To estimate the interstellar extinction $A_V$, we
compared the observed value of $(B-V)$ for the system with the intrinsic
$(B-V)_{\circ}$ resulting from the spectral fit. Then, we computed the
total extinction as $A_V={\cal R}\,E(B-V)$ with ${\cal R}=3.1\pm0.3$. We
obtained values of $E(B-V)=0.19\pm0.03$~mag and $A_V=0.60\pm0.10$~mag. The
overall procedure avoids the use of bolometric corrections and is
self-consistent as it employs the best-fitting model atmospheres (i.e.
with the appropriate $\log g$ and metallicity) to calculate the $B$ and
$V$ magnitudes. The need for spectrophotometry could be circumvented
because we derived $T_{\rm eff}$ from spectra and benefit from the weak
temperature dependency of $(B-V)$ above $\sim$30000 K to obtain an
$E(B-V)$ value (although with a larger error bar).

A basic point is a reliable estimation of the error budget. With the 
uncertainty in $M_V$ accounting for the full correlations of the 
intervening parameters, the rest of the quantities in the distance modulus 
equation are essentially uncorrelated. Thus, the contributions from the 
observed $V$ magnitude, $M_V$, and $A_V$ can be combined quadratically. 
From the parameters in Table \ref{tab:prop} our calculation of the 
distance modulus to this M31 EB results in a value of 
$(m-M)_{\circ}=24.44\pm0.12$ mag or, equivalently, $d=772\pm44$ kpc.  
This distance also corresponds to the center of M31 itself because the 
correction due to the location of the EB is negligible ($\sim$0.3\%). Note 
that the error bar accounts for the random uncertainties of the parameters 
but does not include a possible systematic contribution from the 
atmosphere models. However, we expect such systematic error to be no 
larger than a few percent in flux and therefore to have an effect below 
0.05 mag in distance modulus.

Photometric observations carried out by the {\sc direct} and our own 
surveys have uncovered hundreds of EBs in M31, of which 10--20 are 
suitable for high-precision spectroscopic work (Vilardell et al. 2005). 
Study of such EBs is underway and distance determinations for several new 
systems will occur in the coming years. An improved reddening/temperature 
estimation as well as the analysis of several EB systems will result in a 
direct determination of the M31 distance accurate to better than 5\% or 
0.1 mag in distance modulus.

\acknowledgments

Thanks are due to P. Hadrava, T. Harries, P. Wo\'zniak, and S. Zucker for 
making their {\sc korel}, {\sc tangle}, DIA, and {\sc todcor} codes 
available to us. We are grateful to the Gemini staff for the high-quality 
observations taken in queue mode. This program has been supported by the 
Spanish MCyT grant AyA2003-07736 and NSF/RUI grant no. 0507542. I.~R. 
acknowledges support from the Spanish MEC through a Ram\'on y Cajal 
fellowship.


\begin{deluxetable}{ccrr}
\tablefontsize{\footnotesize}
\tablecaption{Radial velocity measurements
\label{tab:rvs}}
\tablewidth{0pt}
\tablehead{
\colhead{HJD} & 
\colhead{Phase} & 
\colhead{RV$_1$} & 
\colhead{RV$_2$} 
}
\startdata
2453260.90985&0.6278& $-$52.9$\pm$10.8          &$-$382.0$\pm$11.7          \\
2453262.87417&0.1812&$-$338.1$\pm$\phantom{1}7.5&    88.4$\pm$\phantom{1}7.5\\
2453319.87934&0.2403&$-$349.1$\pm$\phantom{1}7.6&   116.4$\pm$\phantom{1}6.2\\
2453321.82033&0.7871&    10.5$\pm$13.9          &$-$426.4$\pm$10.2          \\
2453321.87152&0.8016&  $-$1.1$\pm$\phantom{1}9.3&$-$417.9$\pm$\phantom{1}8.6\\
\enddata
\tablecomments{Units of km~s$^{-1}$}
\end{deluxetable}

\begin{deluxetable}{lcc}
\tablefontsize{\footnotesize}
\tablecaption{Orbital elements and fundamental properties of the system
components 
\label{tab:prop}}
\tablewidth{0pt}
\tablehead{
\colhead{Parameter} & 
\multicolumn{2}{c}{Value} 
}
\startdata
$P$ (d)                            & \multicolumn{2}{c}{3.549694$\pm$0.000010}\\
$T_{\rm min}$ (HJD)                & \multicolumn{2}{c}{2452204.421$\pm$0.003}\\
$i$ ($^{\circ}$)                   & \multicolumn{2}{c}{89.3$\pm$1.8}         \\
$\gamma$ (km s$^{-1}$)             & \multicolumn{2}{c}{$-$173$\pm$4}         \\
$a$ (R$_{\odot}$)                  & \multicolumn{2}{c}{33.0$\pm$0.7}         \\
$q\equiv M_2/M_1$                  & \multicolumn{2}{c}{0.65$\pm$0.03}        \\
${T_{\rm eff}}_2/{T_{\rm eff}}_1 $ & \multicolumn{2}{c}{0.817$\pm$0.015}      \\
$F_2/F_1|_B$\tablenotemark{a}      & \multicolumn{2}{c}{0.487$\pm$0.023}      \\
$F_2/F_1|_V$\tablenotemark{a}      & \multicolumn{2}{c}{0.501$\pm$0.022}      \\
\tableline
                                   &     Primary         &    Secondary       \\
\tableline
$K$\tablenotemark{b} (km s$^{-1}$) &    185$\pm$6        & 285$\pm$6          \\
$r_v\equiv R/a$                    &   0.397$\pm$0.005   &    0.342$\pm$0.005 \\
$M$ (M$_{\odot}$)                  &    23.1$\pm$1.3     &     15.0$\pm$1.1   \\
$R$ (R$_{\odot}$)                  &    13.1$\pm$0.3     &     11.3$\pm$0.3   \\
$\log g$ (cgs)                     &   3.57$\pm$0.03     &    3.51$\pm$0.04   \\
$T_{\rm eff}$ (K)                  &  33,900$\pm$500     &   27,700$\pm$500   \\
$v \sin i$ (km s$^{-1}$)           &    230$\pm$10       &      145$\pm$8     \\
$M_V$ (mag)                        &  $-$5.29$\pm$0.07   &  $-$4.66$\pm$0.07  \\
$(B-V)_{\circ}$                    &  $-$0.28$\pm$0.01   &  $-$0.27$\pm$0.01  \\
\tableline
                                   & \multicolumn{2}{c}{System}               \\
\tableline
$B$\tablenotemark{a} (mag)         & \multicolumn{2}{c}{19.19$\pm$0.02}       \\
$V$\tablenotemark{a} (mag)         & \multicolumn{2}{c}{19.27$\pm$0.02}       \\
$[m/H]$                            & \multicolumn{2}{c}{$-$0.01$\pm$0.06}     \\
$E(B-V)$ (mag)                     & \multicolumn{2}{c}{0.19$\pm$0.03}        \\
$A_V$ (mag)                        & \multicolumn{2}{c}{0.60$\pm$0.10}        \\
$M_V$ (mag)                        & \multicolumn{2}{c}{$-$5.77$\pm$0.06}     \\
$(m-M)_{\circ}$ (mag)              & \multicolumn{2}{c}{24.44$\pm$0.12}       \\
\enddata
\tablenotetext{a}{Out-of-eclipse average: $\Delta \phi=[0.14-0.36]+[0.64-0.86]$}
\tablenotetext{b}{Including non-keplerian corrections}
\end{deluxetable}


\end{document}